\documentclass[preprint,12pt,authoryear]{elsarticle}

\usepackage{amssymb,amsfonts,mathrsfs}
\usepackage{graphicx}
\usepackage{bm}
\usepackage[retainorgcmds]{IEEEtrantools}
\usepackage{amsmath}
\usepackage{subcaption}

\journal{arxiv.org}

\begin{document}

\begin{frontmatter}
\title{Non-equilibrium effects in chaperone-assisted translocation of a stiff polymer }
\author{Rouhollah Haji Abdolvahab}
\address{Physics Department, Iran University of Science and Technology (IUST), 16846-13114, Tehran, Iran.}
\ead{rabdolvahab@gmail.com}

\begin{abstract}
Chaperone assisted biopolymer translocation is the main model proposed for translocation \textit{in vivo}. A dynamical Monte Carlo method is used to simulate the translocation of a stiff homopolymer through a nanopore driven by chaperones. Chaperones are proteins that bind to the polymer near the wall and prevent its backsliding through Cis side. The important parameters include binding energy, size and the local concentration of the chaperones. The profile of these local concentrations, build up the chaperones distribution. Here we investigate the effects of binding energy, size and the exponential distribution of chaperones in their equilibration in each step of the polymer translocation needed for stable translocation time. The simulation results show that in case of chaperones with size of a monomer ($\lambda=1$) and/or positive effective binding energy and/or uniform distribution, the chaperones binding equilibration rate/frequency is less than $5$ times per monomer. However, in some special cases in exponential distribution of chaperones with size $\lambda>1$ and negative effective binding energy the equilibration rate will diverge to more than $20$ times per monomer. We show that this non-equilibrium effect results in supper diffusion, seen before. Moreover, we confirm the equilibration process theoretically.

\end{abstract}


\end{frontmatter}

\section{Introduction}
\label{intro}

Translocation of biomolecules through the nanopores \cite{Meller03} is one of the most important processes within biological cells. This is a ubiquitous process in cell metabolism. Protein's translocation through endoplasmic reticulum is an example. The polymer translocation also seen in proteins transport through organelles like mitochondria \cite{Al02,muthuAnn07,rapaport}. Translocation of messenger RNA through nuclear pore complexes in gene expression and in transcription through eukaryotic cells are two other biological instances \cite{Al02}. Translocation of DNA through protein channels covering the bacterial membrane amid phage infection is another example \cite{03mellerBJ,94Dreiseikelmann}. Biotechnological examples also includes gene therapy, drug delivery and cheap rapid sequencing of the biopolymers \cite{Marzio03,Nakane03,Branton08,Ramin12,Fanzio12,Wanunu15,Liang15}. The experimental work of Kasianowicz \textit{et al.} \cite{Kasianowicz96} on ssRNA translocation through an $\alpha$-hemolysin channel was an influential work. Hereafter, there has been many experimental and theoretical works and simulations in polymer translocation \cite{Meller03,Panja13,Sun14,metzr14}.

There are many different mechanisms to drive translocation of polymers. \textit{In vitro}, people usually use a strong electric field to drive the translocation of highly charged biopolymers like single stranded DNA or RNA. Moreover, there could be many other parameters affecting the translocation such as crowding \cite{Gopinathan07,16PuJCP}, pressure and confinement \cite{07Molineux,10PanjaPB,13PanjaNRM,metzr14}. However, the most important model for the translocation \textit{in vivo} is chaperone assisted translocation \cite{Tomkiewicz07}. This model with the name of Brownian ratchet mechanism was first proposed by Simon \textit{et al.} in 1992 \cite{Si&Pe&Os}. In this model proteins called chaperones are bound to the polymer in the Trans side and actively pull the polymer or just prevent its backsliding through the Cis side \cite{Al02,Tomkiewicz07}. Later, the experiments of Matlack \textit{et al.} in 1999 highlights the problem again \cite{L&R&H,elston02,Z&R}. Subsequently, many theoretical and simulations struggled to have a better understanding of different aspects of the problem \cite{metzr14,A&M4,Abd08,AbdPRE11,AbdJCP11,kaifujacs,kaifuPRE14,Cao15,Suhonen16,AbdPLA16}.

There are many works on the non-equilibrium aspects of forced polymer translocation \cite{07SakauePRE,10BhattacharyaPRE,13SakauePRE,16HaanJCP}. This out of equilibrium property emerge as a result of force pulling the polymer through the Trans side. Here we investigate the chaperone assisted translocation which is used \textit{in vivo}. Following our recent works we consider the chaperones exponential distribution effects on polymer translocation \cite{AbdPLA16, AbdEPJ17}. We will show that even in the case of stiff polymer the distribution may induce the non-equilibrium effects in the translocation process.

In what follows we examine the chaperones binding rate effects on the translocation. Hereafter introducing our dynamical Monte Carlo simulation, we will discuss our simulation results in our main part of article in section \ref{result}. Finally we will sum up our findings in the conclusion.

\section{Simulation}
\label{sim}

\subsection{Theoretical model}
\label{simod}

\begin{figure}
\includegraphics[width=0.75\textwidth]{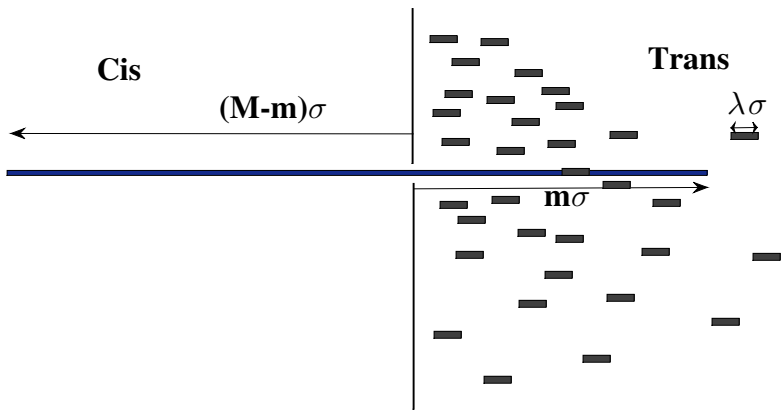}
\centering
\caption{A stiff polymer is translocating from the Cis side (left) to the Trans side (right). Chaperones of size $\lambda\sigma$ are distributed in the Trans side. The total length of the polymer is $L=M\sigma$. $m$ monomers are no translocated to the right.}
\label{Model}
\end{figure}

As the figure \ref{Model} shows, we simulate a stiff homopolymer consisted of $M$ monomer with size of $L=M\sigma$. Chaperones with the same size of $\lambda\sigma$ are distributed only in the Trans side. We suppose the $\lambda$ to be an integer \cite{AbdPRE11,A&M4}. They have local concentration which depends on their distribution. Moreover, there is a binding energy, positive or negative, between the chaperones and the polymer.

They bind (unbind) to (from) the polymer in the Trans side. The polymer always can go to the right. However, for backsliding of the polymer, its near the wall site must be unbound. The wall has no width \cite{AbdEPJ17}. Binding of chaperones bias the translocation through the Trans side. The master equation for this process is written as:

\begin{eqnarray}
\label{me}
\frac{\partial P(m,t)}{\partial t} &=& \mathcal{W}^+(m-1)P(m-1,t-1)\cr
\nonumber \\&+&\mathcal{W}^-(m+1)P(m+1,t+1)\cr
\nonumber \\&-&(\mathcal{W}^+(m)+\mathcal{W}^-(m))P(m,t)
\end{eqnarray}

 in which $\mathcal{W}^\pm$ are the transfer rates for translocating polymer to the right and left. $P(m,t)$ is the probability of finding polymer in time $t$ at condition in which $m$ monomer of it translocated to the right. Using transfer rates,  $\mathcal{W}^\pm$, and boundary conditions one could find the translocation time by calculating its mean first passage time \cite{Ga}.

\subsection{Describing the Monte-Carlo method}
\label{mc}

We use a dynamical Monte Carlo method to simulate the translocation of a stiff polymer as follows. The polymer always can go to the right with probability of half. However, backsliding of the polymer is restricted. The polymer may come back through the Cis side only if there is not any chaperone bound to the polymer near the wall. Moreover, we use from the so called transmission boundary condition \cite{redner}. It means reflective at first and absorbing at the end. As a result the polymer does not come back to the Cis when the first monomer is near the wall \cite{AbdEPJ17}. Chaperones will try to bind/unbind in each step of the translocation by frequency $f$ per monomer per $40$. It means, for example, in the case of a polymer with $m=40$ and by the frequency $f=40$, the binding/unbinding process is one time. A monomer in the Trans side is selected randomly. If there is a chaperone bound to it we try to unbind it with its probability and \textit{vice versa} (it will try to bind a chaperone accordingly). Due to our computational limits, we changed the frequency from $1$ to $10^3$.

\textit{Chaperones binding probability:} There are three terms in the binding probability. Boltzmann distribution, which depend on the binding energy between chaperones and the polymer. Entropy linked to different patterns in which chaperones may distributed on the polymer and availability of the chaperones related to its local density \cite{A&M4,AbdPRE11}. The second term is automatically comes in the simulation. In place of the binding energy, we define effective binding energy (EBE or $\mathscr{E}_{eff}$) to combine the first and third term as \cite{AbdPRE11} $\mathscr{E}_{eff}\equiv-\frac{1}{\lambda}\log\left[c_0v_0\exp\left(-
\varepsilon/k_BT\right)\right]$, where $\varepsilon$ comes for the chaperone binding energy per monomer of the polymer. $c_0$ denotes the chaperone concentration, and $v_0$ stands for their volume \cite{AbdJCP11,AbdEPJ17}. Thus the binding and unbinding probabilities are written as:

\begin{eqnarray}
P_{bind}   =\frac{\exp\left(-\sum_{i=1}^\lambda \mathscr{E}_{eff}^i\right)}{1+\exp\left(-\sum_{i=1}^\lambda \mathscr{E}_{eff}^i\right)},
P_{unbind} = \frac{1}{1+\exp\left(-\sum_{i=1}^\lambda \mathscr{E}_{eff}^i\right)}.
\label{pr}
\end{eqnarray}

The effective binding energy is changed from $-4$ to $4$ and the polymer length is restricted to $M=50$. In order to reach to an acceptable error, we repeat the translocation process for at least $10^4$ times. Moreover, we use the chaperones of different sizes of $\lambda=1, 2$ and $6$.

\textit{Chaperones distributions:} For simplicity we restrict the chaperones to distributed only in the right part. We consider the exponential distribution with different rates, $\alpha$, for the chaperones and compare its results with the usual uniform distribution ($\alpha=0$). In order to change the chaperones distribution in our Monte Carlo simulation it is enough to change the $\mathscr{E}_{eff}$ to $\mathscr{E}_{eff} + \alpha d$ in which $d$ is the distance (per monomer size) between the wall and the monomer in which we need its near chaperones concentration \cite{AbdPLA16,AbdEPJ17}.

\section{Results and discussion}
\label{result}

\subsection{Mean translocation time}
\label{enl}

\begin{figure}
\centering
    \begin{subfigure}[b]{0.95\textwidth}
    \includegraphics[width=1\linewidth]{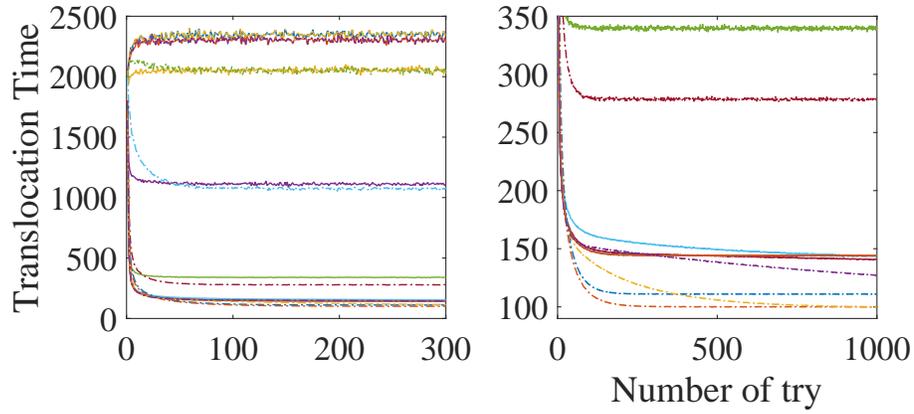}
    \caption{}
    \label{Tr}
    \end{subfigure}

    \begin{subfigure}[b]{0.95\textwidth}
    \includegraphics[width=1\linewidth]{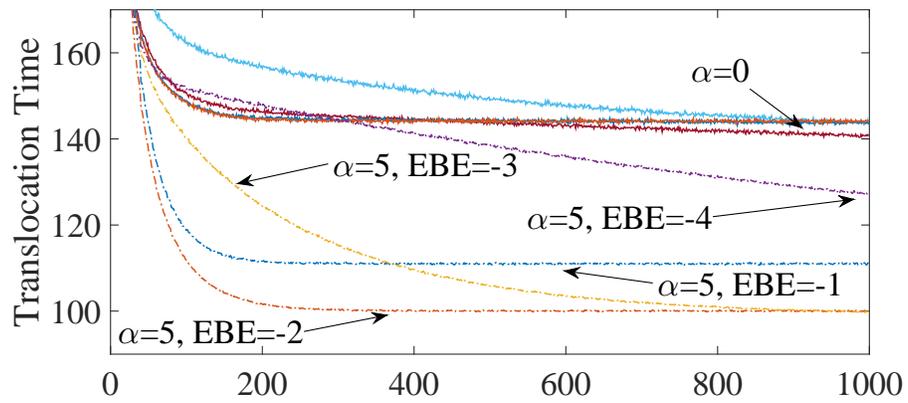}
    \caption{}
    \label{trz}
    \end{subfigure}
\caption{Translocation time of polymers, constructed of 50 monomers, versus try number. Chaperones size are $\lambda=2$ and exponential rates are $\alpha=0, 5$. The dash-dotted lines are for $\alpha=5$ while the solid lines stand for  $\alpha=0$. The right figure is a zoom on the left one.}
\label{tr0}
\end{figure}

We simulate 1 dimensional stiff homopolymer translocation through a nanopore using a dynamic Monte Carlo method in presence of the chaperones with different sizes and different spatial distributions. There are chaperones with distinct spatial distributions and various EBEs by monomers in the Trans side(There is not any chaperone in the Cis side.). Try number, or chaperones rate of binding (its frequency denotes by $f$), is an important parameter in calculating the translocation time of the polymer. In spite of this importance, there are few works on investigating its effects on the polymer translocation \cite{07Orsogna}. People suppose that due to interaction of the polymer with the pore and its size, the chaperones will reach to equilibrium in each step of the polymer translocation \cite{A&M4}. In what follows we will show that although this is true for the case of uniform distribution of chaperones, in exponential distribution, the equilibration frequency will become large and the assumption is violated.

We translocate polymers in presence of chaperones of sizes $\lambda=1, 2, 6$ and spatial distributions of $\alpha=0, 1, 5, 10$ and $EBE=-4:+4$. The figure \ref{tr0} shows mean translocation time versus frequency or rate of the chaperones for different exponential chaperone distributions of $\alpha=0, 5$ and different EBEs for chaperones of size $\lambda=2$. Different curves stand for different chaperones spatial distribution and/or different EBEs. The right figure is a zoom of the left one for faster polymers. Note that $\alpha=0$ means uniform spatial distribution of the chaperones.

The simulation results show that there are different regimes based on convexity and/or equilibrium rate:
\begin{itemize}
  \item In large enough and positive EBEs the time versus rate curves are strictly ascending. They soon will reach to their equilibrium rates.
  \item In large enough and negative EBEs the time versus rate curves are strictly descending. They quickly will reach to their equilibrium rates usually but in some special cases the equilibrium rate will become quite large.
  \item In some intermediate energies, we will see a maximum in time versus rate curves.
\end{itemize}

\begin{figure}
\includegraphics[width=12cm]{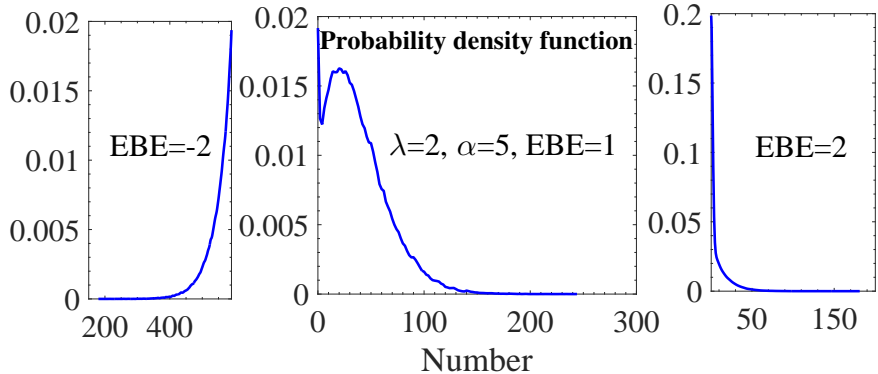}
\centering
\caption{Probability distribution function of finding the near the wall monomer of the polymer bound is compared for chaperones of size $\lambda=2$ and exponential distribution with $\alpha=5$ for three different $\mathscr{E}_{eff}=-2, 1, 2$.}
\label{bpdf}
\end{figure}

In large positive EBEs, the chaperones do not prefer to bind to the polymer. Hence, the probability density function of finding the monomers bound are well asymmetric to the left (see the figure \ref{bpdf}). In asymmetric density functions of this kind, mean is always in right and close to the most probable. We will show that in this situation increasing the fluctuation, will increase the $P_{bind}$ (near the wall binding probability). Consequently, by increasing the rate and decreasing the fluctuation, the translocation time will be increased (see the \ref{app11}). In contrast in large negative EBEs, increasing the frequency will cause the translocation time to be decreased (see the \ref{app12}).

\subsection{Equilibrium rate}
\label{seqr}

\begin{figure}
\includegraphics[width=12cm]{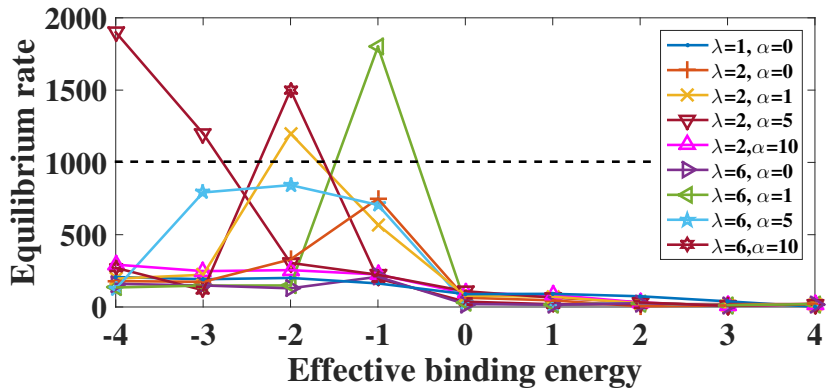}
\centering
\caption{Equilibrium rates are plotted against EBEs for different binding sizes $\lambda=1, 2, 6$ and different chaperones distributions of $\alpha=0, 1, 5 and 10$. The dashed line shows the maximum rate in the simulation. The points over the black dashed line are not real data and just estimated (based on figure \ref{tr0}).}
\label{eqr}
\end{figure}

As discussed in the previous section, an important parameter in describing the polymer translocation is the equilibrium rate (the rate from which the translocation time does not change). In this regard, the polymers have two different behaviours. All the polymers with positive $\mathscr{E}_{eff}$, which means the chaperones prefer not to bind to the polymer, have the small equilibrium rates. In contrast in the case of negative $\mathscr{E}_{eff}$, the equilibrium rate could be relatively large. Indeed, our simulation results show that this non-equilibrium properties comes from the chaperones exponential distribution. We did not see any divergence in the uniform distribution of the chaperones (compare the plots in figure \ref{eqr}). The equilibrium rate of the polymers translocation in vicinity of the chaperones with different sizes of $\lambda=1, 2, 6$ and with chaperones spatial distribution with different exponential rates of $\alpha=0, 1, 5, 10$ are plotted against $\mathscr{E}_{eff}$ in figure \ref{eqr}. As it shows there is not any important divergence in the cases of uniform distribution, $\alpha=0$, and/or chaperones with the size of a monomer, $\lambda=1$. To better understand this results one should find a more detailed description of the chaperones exponential distributions effects on the polymer translocation (see \cite{AbdEPJ17}).

It is pertinent to mention that our results paved the way for understanding the supper diffusion reported in our previous work \cite{AbdEPJ17}. In that article we show that in some special case the scaling exponent of time \textit{vs} polymer length, $\beta$ ($T\sim M^\beta$), becomes less than 1 ($\beta<1$). It can be explained as follows. Increasing the polymer length will increase the time for chaperones to equilibrate. Consequently, the equilibration rate will be decreased. Hence, trying to calculate the scaling exponent $\beta$ before equilibrium will cause the translocation time increase less than the expected amount and the exponents could even become less than $1$.

\subsection{Mean waiting time}
\label{mwt}

\begin{figure}
\centering
    \begin{subfigure}[b]{0.95\textwidth}
    \includegraphics[width=1\linewidth]{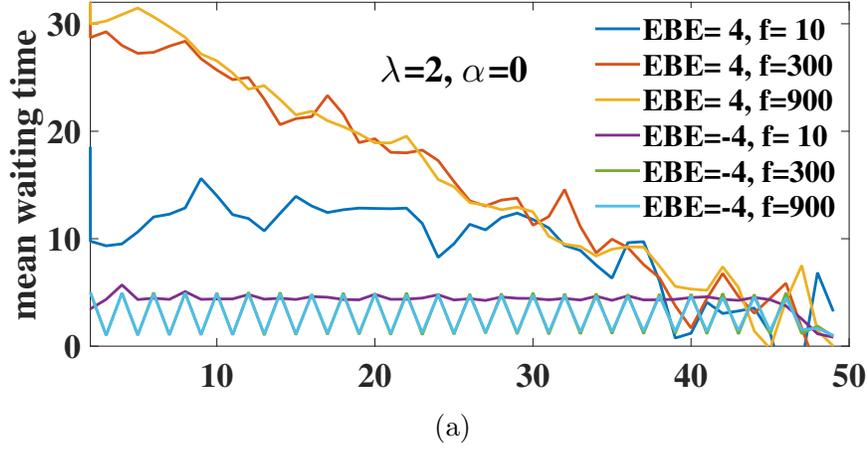}
    \caption{}
    \label{mw1}
    \end{subfigure}

    \begin{subfigure}[b]{0.95\textwidth}
    \includegraphics[width=1\linewidth]{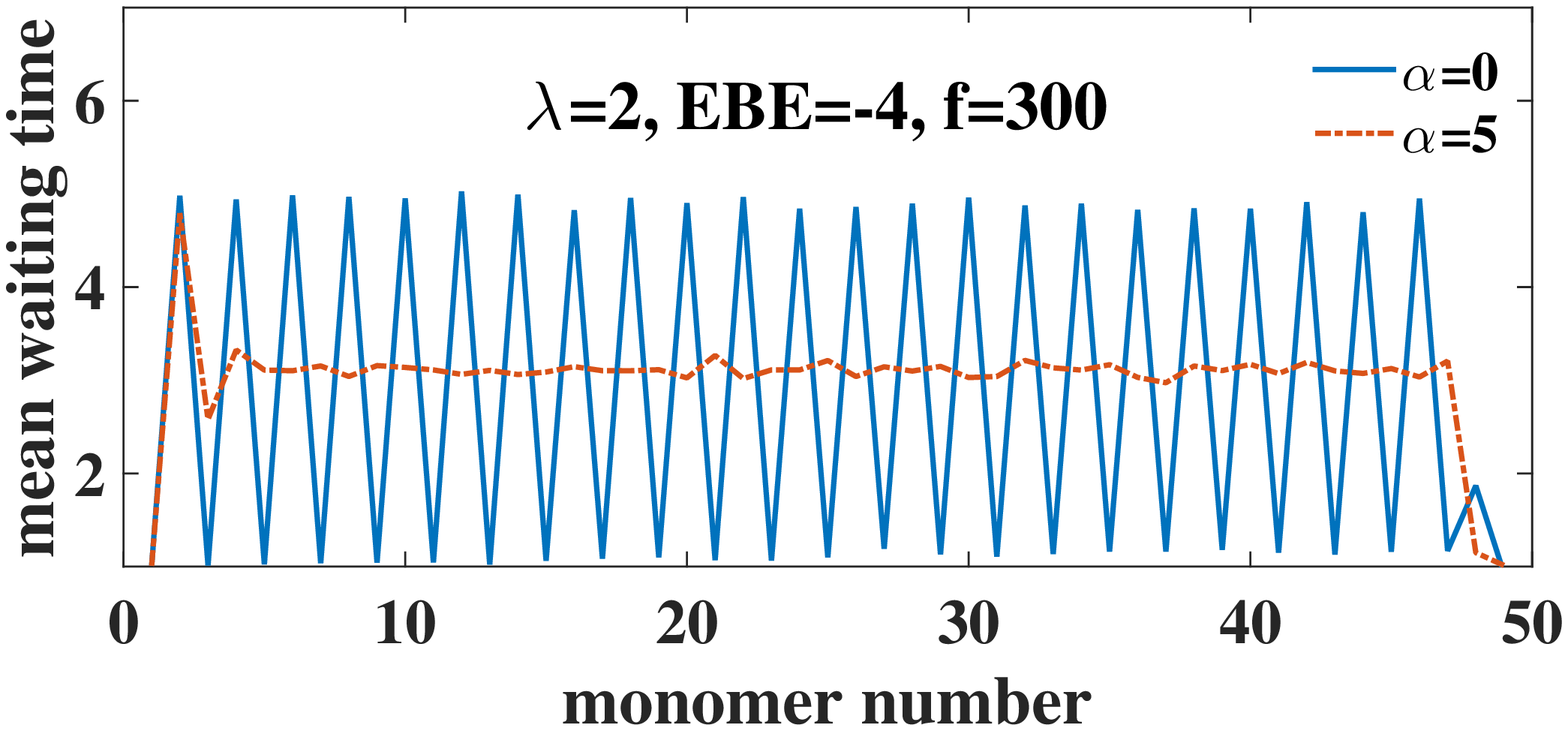}
    \caption{}
    \label{mw2}
    \end{subfigure}
\caption{Mean waiting times of polymer translocation in different conditions are compared. In above figure, the chaperones have uniform distribution and different EBEs of $\mathscr{E}_{eff}=-4, 4$ and rate frequencies of $f=10, 300, 900$ are compared(note that the chaperones with $\mathscr{E}_{eff}=-4$ and rate frequencies of $f=300, 900$ are exactly the same.). In the figure \ref{mw2}, we compare the uniform and exponential distribution of chaperones.}
\label{mw}
\end{figure}

Details of the translocation can be seen from its waiting times. Mean Waiting Time (MWT) of the translocation for chaperones of size $\lambda=2$ and $\mathscr{E}_{eff}=-4, 4$ for uniform chaperone's distribution and $3$ different rates of $f=10, 300, 900$ is shown in figure \ref{mw1}. As it shows MWT for $\mathscr{E}_{eff}=4$ and for rates of $f\geq 300$ are decreasing linearly. However, for the same $\mathscr{E}_{eff}$ but for smaller rate of $f=10$, it starts from smaller amounts and it makes its translocation velocity faster. In contrast, in case of $\mathscr{E}_{eff}=-4$, the translocation velocity increase by increasing the rate from $f=10$ to $f=300, 900$. As a result of the large binding probability the MWT for the rates $f=300, 900$ becomes completely sawtooth shape.

As we saw in figure \ref{trz}, the translocation time for two different case of uniform distribution and exponential distribution with $\alpha=5$, both with chaperones of size $\lambda=2$ and $\mathscr{E}_{eff}=-4$ in rates of $f=300$ equalize to each other. MWT for these two case is plotted in figure \ref{mw2}. Because of the large negative $\mathscr{E}_{eff}$, the chaperones bind to the polymer as right as it find a free place near the wall. In the case of uniform distribution availability of the chaperones are the same through the whole polymer which cause the sawtooth shape.However, in the case of exponential distribution, the chaperones may unbind from the polymer in further sites away from the wall. Consequently, the sawtooth shape becomes smooth or disappeared. On the other hand, in this case it takes time for the system to find its equilibrium and thus we see the translocation time is decreased by rate frequency and in $f=300$ it coincide with its uniform counterparts.

\section{Conclusions}
\label{conc}

We simulate the translocation of stiff homopolymer through a nanopore driven by chaperones. We investigate specially the chaperones binding frequency and spatial distribution on the translocation time. Our results show that there are different patterns of equilibration in terms of chaperones size, effective binding energy and spatial distribution. In most cases the equilibrium is reaching soon (less than $10$ try per monomer).  However, an increase in equilibration frequency (more than $20$ try per monomer) is seen in the cases of $\lambda>1$, $\mathscr{E}_{eff}<0$ and $\alpha>0$ roughly in the interval of $-6<\lambda \mathscr{E}_{eff}+\alpha<-1$. In larger amounts the chaperones do not prefer to bind and in less amounts the chaperones could not unbound from further sites. This result pave the way for understanding the supper diffusion reported in \cite{AbdEPJ17}. We also strengthen our simulation results by theoretical discussion about the effect of chaperones binding rate on translocation time.

\begin{appendix}

\section{Fluctuation in binding probability and translocation time}
\label{app1}

Based on a master equation approach and using mean first passage time theory, we provide translocation time as follows \cite{AbdPRE11}:

\begin{equation}\label{tp}
  T=\frac{2\tau_0}{P_{bind}}\left(N+1-\frac{1-P_{bind}}{P_{bind}}\left[1-(1-P_{bind})^{N+1}\right] \right),
\end{equation}

where $\tau_0$ is the time takes for a bare polymer to translocate over distance of a monomer, $N$ is the total number of the monomers and $P_{bind}$ is the probability of the polymer to be bound near the wall. In the small and large amount of $P_{bind}$ one may approximate the equation \ref{tp} as follows:

\subsection{Large binding probability}
\label{app11}

In large enough binding probability $P_{bind}\simeq 1$ which occurs in large negative EBE, expansion of the equation \ref{tp} leads to:

\begin{equation}\label{tlp}
  T\simeq \frac{2\tau_0}{P_{bind}}(N+1).
\end{equation}

Fluctuation of $P_{bind}$ which shows itself in translocation time can be investigated as follows. Admit the change in $P_{bind}$ to be of order $\delta$; $P_{bind}\rightarrow P_{bind}\pm \delta$. We then average over this range by integration:

\begin{equation}\label{tlpf}
  \overline{T}_{large}\simeq\frac{1}{2\delta}\left(\int_{P_{bind}-\delta}^{P_{bind}+\delta}\frac{2\tau_0}{P}(N+1)dP\right)\simeq\frac{2\tau_0(N+1)}{P_{bind}-\delta},
\end{equation}

where $\overline{T}_{large}$ is average of $T$ over fluctuations of $P_{bind}$ for large binding probabilities. As expected from our simulation results, it shows that increasing the chaperones binding rate which decreases the fluctuation $\delta$, will reduce the mean translocation time.

\subsection{Small binding probability}
\label{app12}

In contrast to the previous section, in small binding probability, expansion of equation \ref{tp} results in:

\begin{eqnarray}\label{tsp}
&&T(P_{bind}) = 2\tau_0(N + 1) \times \left(C_0(N) - C_1(N)P_{bind}+C_2(N)P_{bind}^2- \cdots\right),\\
&&C_0(N)=1 + \frac{N}{2}, C_1(N)=\frac{N}{2}\left(1 + \frac{N-1}{3}\right), C_2(N)=\frac{N(N-1)}{3!}\left(1+\frac{N-2}{4}\right).\nonumber
\end{eqnarray}

Note that this approximation is true when $NP_{bind}\ll 1$ and the important parameter here is the P\'{e}clet number not $P_{bind}$ itself (see \cite{AbdJCP11} for more detail). As the figure \ref{bpdf} shows, the binding probability distribution is exponential. Let assume its exponential determines by parameter $a$; $P(P_{bind})\propto \exp(-aP_{bind})$. Averaging of the translocation time over fluctuations of $P_{bind}$ leads to:

\begin{eqnarray}\label{tspf}
 &&Z\equiv \left(\int_{P_{bind}-\delta}^{P_{bind}+\delta}\exp(-aP)dP\right)\\\nonumber
 &&\Rightarrow \overline{P}_{bind}=-\frac{\partial}{\partial a}\ln(Z) \simeq -\frac{\partial}{\partial a}\ln\left(2\delta \exp(-aP_{bind})(1-(a\delta)^2)\right) \simeq P_{bind}+2a\delta^2\\\nonumber
 &&\overline{T}_{small}-T=\frac{1}{2\delta}\left(\int_{P_{bind}-\delta}^{P_{bind}+\delta}T(P)\exp(-aP)dP\right)-T(P_{bind})\\
&&\Rightarrow \overline{T}_{small}-T \simeq -C_1(N)2a\delta^2.
\end{eqnarray}

Increasing the chaperones binding rate will decrease the fluctuation over $P_{bind}$, $\delta$, and as a result the mean translocation time, $\overline{T}_{small}$, will be increased. It is in place to note here that the $EBE$ in which the regimes changed, say $EBE_0$, may be obtained by comparison of second and third term in equation \ref{tsp}:

\begin{equation}\label{rcc}
C_1(N)P_{bind}>C_2(N)P_{bind}^2\Rightarrow \frac{N}{3}P_{bind}<1.
\end{equation}

Presume the binding probability to be proportional to its Boltzmann distribution ($P_{bind}\propto \exp(-\lambda EBE)$) leads us to $EBE_0\simeq \frac{1}{\lambda}\ln(\frac{N}{3})$. As an example in case of $\lambda=2$ and $N=50$, $EBE_0\simeq 1.4$ which is compatible with the simulation results (see \textit{e.g.} figure \ref{Tr}).

\end{appendix}


\bibliographystyle{elsarticle-harv}
\section*{\refname}
\bibliography{MyReferencesNN960418}

\begin{thebibliography}{44}
\expandafter\ifx\csname natexlab\endcsname\relax\def\natexlab#1{#1}\fi
\expandafter\ifx\csname url\endcsname\relax
  \def\url#1{\texttt{#1}}\fi
\expandafter\ifx\csname urlprefix\endcsname\relax\def\urlprefix{URL }\fi

\bibitem[{Abdolvahab(2016)}]{AbdPLA16}
Abdolvahab, R.~H., 2016. Investigating binding particles distribution effects
  on polymer translocation through nanopore. Physics Letters A 380,
  1023–1030.

\bibitem[{Abdolvahab(2017)}]{AbdEPJ17}
Abdolvahab, R.~H., 2017. Chaperone driven polymer translocation through
  nanopore: spatial distribution and binding energy. The European Physical
  Journal E 40, 41.

\bibitem[{Abdolvahab et~al.(2011{\natexlab{a}})Abdolvahab, Ejtehadi, and
  Metzler}]{AbdJCP11}
Abdolvahab, R.~H., Ejtehadi, M.~R., Metzler, R., 2011{\natexlab{a}}. First
  passage time distribution of chaperone driven polymer translocation through a
  nanopore: Homopolymer and heteropolymer cases. Journal Of Chemical Physics
  135, 5102.

\bibitem[{Abdolvahab et~al.(2011{\natexlab{b}})Abdolvahab, Ejtehadi, and
  Metzler}]{AbdPRE11}
Abdolvahab, R.~H., Ejtehadi, M.~R., Metzler, R., 2011{\natexlab{b}}. Sequence
  dependence of the binding energy in chaperone-driven polymer translocation
  through a nanopore. Physical Review E 83, 011902.

\bibitem[{Abdolvahab et~al.(2008)Abdolvahab, Roshani, Nourmohammad, Sahimi, and
  Tabar}]{Abd08}
Abdolvahab, R.~H., Roshani, F., Nourmohammad, A., Sahimi, M., Tabar, M. R.~R.,
  2008. Analytical and numerical studies of sequence dependence of passage
  times for translocation of heterobiopolymers through nanopores. Journal of
  Chemical Physics 129~(235102), 1--8.

\bibitem[{Alberts et~al.(2002)Alberts, Johnson, Lewis, Raff, Roberts, and
  Walter}]{Al02}
Alberts, B., Johnson, A., Lewis, J., Raff, M., Roberts, K., Walter, P., 2002.
  Molecular Biology of the Cell. Garland Publishing, New York.

\bibitem[{Ambjörnsson and Metzler(2004)}]{A&M4}
Ambjörnsson, T., Metzler, R., 2004. Chaperone-assisted translocation. Physical
  Biology 1, 77.

\bibitem[{Bates et~al.(2003)Bates, Burns, and Meller}]{03mellerBJ}
Bates, M., Burns, M., Meller, A., 2003. Dynamics of dna molecules in a membrane
  channel probed by active control techniques. Biophysical journal 84~(4),
  2366--2372.

\bibitem[{Bhattacharya and Binder(2010)}]{10BhattacharyaPRE}
Bhattacharya, A., Binder, K., 2010. Out-of-equilibrium characteristics of a
  forced translocating chain through a nanopore. Physical Review E 81~(4),
  041804.

\bibitem[{Branton et~al.(2008)Branton, Branton, Deamer, Marziali, Bayley,
  Benner, Butler, Ventra, Garaj, Hibbs, Huang, Jovanovich, Krstic, Lindsay,
  Ling, Mastrangelo, Meller, Oliver, Pershin, Ramsey, Riehn, Soni,
  Tabard-Cossa, Wanunu, Wiggin, Schloss, Deamer, Marziali, Bayley, Benner,
  Butler, Ventra, Garaj, Hibbs, Huang, Jovanovich, Krsticand, Lindsay, Ling,
  Mastrangelo, Meller, Oliver, Pershin, Ramsey, Riehn, Soni, Tabard-Cossa,
  Wanunu, Wiggin, and Schloss}]{Branton08}
Branton, D., Branton, D., Deamer, D.~W., Marziali, A., Bayley, H., Benner,
  S.~A., Butler, T., Ventra, M.~D., Garaj, S., Hibbs, A., Huang, X.,
  Jovanovich, S.~B., Krstic, P.~S., Lindsay, S., Ling, X.~S., Mastrangelo,
  C.~H., Meller, A., Oliver, J.~S., Pershin, Y.~V., Ramsey, J.~M., Riehn, R.,
  Soni, G.~V., Tabard-Cossa, V., Wanunu, M., Wiggin, M., Schloss, J.~A.,
  Deamer, D.~W., Marziali, A., Bayley, H., Benner, S.~A., Butler, T., Ventra,
  M.~D., Garaj, S., Hibbs, A., Huang, X., Jovanovich, S.~B., Krsticand, P.~S.,
  Lindsay, S., Ling, X.~S., Mastrangelo, C.~H., Meller, A., Oliver, J.~S.,
  Pershin, Y.~V., Ramsey, J.~M., Riehn, R., Soni, G.~V., Tabard-Cossa, V.,
  Wanunu, M., Wiggin, M., Schloss, J.~A., 2008. The potential and challenges of
  nanopore sequencing. Nature Biotechnology 20, 1146.

\bibitem[{Carson and Wanunu(2015)}]{Wanunu15}
Carson, S., Wanunu, M., 2015. Modulating dna translocation by a controlled
  deformation of a pdms nanochannel device. Nanotechnology 26, 074004.

\bibitem[{Cohen et~al.(2012)Cohen, Chaudhuri, and Golestanian}]{Ramin12}
Cohen, J.~A., Chaudhuri, A., Golestanian, R., 2012. Stochastic sensing of
  polynucleotides using patterned nanopores. Physical Review X 2, 2160.

\bibitem[{D'Orsogna et~al.(2007)D'Orsogna, Chou, and Antal}]{07Orsogna}
D'Orsogna, M.~R., Chou, T., Antal, T., 2007. Exact steady-state velocity of
  ratchets driven by random sequential adsorption. Journal of Physics A:
  Mathematical and Theoretical 40~(21), 5575.

\bibitem[{Dreiseikelmann(1994)}]{94Dreiseikelmann}
Dreiseikelmann, B., 1994. Translocation of dna across bacterial membranes.
  Microbiological reviews 58~(3), 293--316.

\bibitem[{Elston(2002)}]{elston02}
Elston, T.~C., 2002. The brownian ratchet and power stroke models for
  posttranslational protein translocation into the endoplasmic reticulum.
  Biophysical Journal 82, 1239.

\bibitem[{Fanzio et~al.(2012)Fanzio, Manneschi, Angeli, Mussi, Firpo,
  Ceseracciu, Repetto, and Valbusa}]{Fanzio12}
Fanzio, P., Manneschi, C., Angeli, E., Mussi, V., Firpo, G., Ceseracciu, L.,
  Repetto, L., Valbusa, U., 2012. Modulating dna translocation by a controlled
  deformation of a pdms nanochannel device. Scientific Report 2, 791.

\bibitem[{Gardiner(2002)}]{Ga}
Gardiner, C.~W., 2002. Handbook of Stochastic Methods for Physics, Chemistry
  and the Natural Sciences, 2nd Edition. Vol.~13 of Synergetics. Springer, New
  York.

\bibitem[{Gopinathan and Kim(2007)}]{Gopinathan07}
Gopinathan, A., Kim, Y.~W., 2007. Polymer translocation in crowded
  environments. Physical Review Letters 99, 228106.

\bibitem[{Grayson and Molineux(2007)}]{07Molineux}
Grayson, P., Molineux, I.~J., 2007. Is phage dna ‘injected’into
  cells—biologists and physicists can agree. Current opinion in microbiology
  10~(4), 401--409.

\bibitem[{Kasianowicz et~al.(1996)Kasianowicz, Brandin, Branton, and
  Deamer}]{Kasianowicz96}
Kasianowicz, J.~J., Brandin, E., Branton, D., Deamer, D.~W., 1996.
  Characterization of individual polynucleotide molecules using a membrane
  channel. Proceedings of the National Academy of Sciences U.S.A. 93, 13770.

\bibitem[{Liang and Zhang(2015)}]{Liang15}
Liang, F., Zhang, P., 2015. Nanopore dna sequencing: Are we there yet? Science
  Bulletin 60, 296.

\bibitem[{Liebermeister et~al.(2001)Liebermeister, Rapoport, and
  Heinrich}]{L&R&H}
Liebermeister, W., Rapoport, T.~A., Heinrich, R., 2001. Ratcheting in
  post-translational protein translocation: a mathematical model. Journal of
  Molecular Biology 305, 643.

\bibitem[{Marzio and Kasianowicz(2003)}]{Marzio03}
Marzio, E.~D., Kasianowicz, J.~J., 2003. Phase transitions within the isolated
  polymer molecule: Coupling of the polymer threading a membrane transition to
  the helix-random coil, the collapse, the adsorption, and the equilibrium
  polymerization transitions. Journal of Chemical Physics 119, 6378.

\bibitem[{Meller(2003)}]{Meller03}
Meller, A., 4 2003. Dynamics of polynucleotide transport through
  nanometre-scale pores. Journal of Physics: Condensed Matter 15, R581.

\bibitem[{Molineux and Panja(2013)}]{13PanjaNRM}
Molineux, I.~J., Panja, D., 2013. Popping the cork: mechanisms of phage genome
  ejection. Nature reviews. Microbiology 11~(3), 194.

\bibitem[{Muthukumar(2007)}]{muthuAnn07}
Muthukumar, M., 2007. Mechanism of dna transport through pores. Annual Review
  of Biophysics and Biomolecular Structure 36, 435.

\bibitem[{Nakane et~al.(2003)Nakane, Akeson, and Marziali}]{Nakane03}
Nakane, J.~J., Akeson, M., Marziali, A., 2003. Nanopore sensors for nucleic
  acid analysis. Journal of Physics: Condensed Matter 15~(32), R1365.

\bibitem[{Palyulin et~al.(2014)Palyulin, Ala-Nissila, and Metzler}]{metzr14}
Palyulin, V.~V., Ala-Nissila, T., Metzler, R., 2014. Modulating dna
  translocation by a controlled deformation of a pdms nanochannel device. Soft
  Matter 10, 9016.

\bibitem[{Panja et~al.(2013)Panja, Barkema, and Kolomeisky}]{Panja13}
Panja, D., Barkema, G.~T., Kolomeisky, A.~B., 2013. Through the eye of the
  needle: recent advances in understanding biopolymer translocation. Journal of
  Physics: Condensed Matter 25~(41), 413101.

\bibitem[{Panja and Molineux(2010)}]{10PanjaPB}
Panja, D., Molineux, I.~J., 2010. Dynamics of bacteriophage genome ejection in
  vitro and in vivo. Physical biology 7~(4), 045006.

\bibitem[{Pu et~al.(2016)Pu, Jiang, and Hou}]{16PuJCP}
Pu, M., Jiang, H., Hou, Z., 2016. Polymer translocation through nanopore into
  active bath. The Journal of chemical physics 145~(17), 174902.

\bibitem[{Rapoport(2007)}]{rapaport}
Rapoport, T.~A., 2007. Protein translocation across the eukaryotic endoplasmic
  reticulum and bacterial plasma membranes. Nature 450, 663.

\bibitem[{Redner(2001)}]{redner}
Redner, S., 2001. A guide to first-passage processes. Cambridge University
  Press, Cambridge UK.

\bibitem[{Saito and Sakaue(2013)}]{13SakauePRE}
Saito, T., Sakaue, T., 2013. Cis-trans dynamical asymmetry in driven polymer
  translocation. Physical Review E 88~(4), 042606.

\bibitem[{Sakaue(2007)}]{07SakauePRE}
Sakaue, T., 2007. Nonequilibrium dynamics of polymer translocation and
  straightening. Physical Review E 76~(2), 021803.

\bibitem[{Simon et~al.(1992)Simon, Peskin, and Oster}]{Si&Pe&Os}
Simon, S.~F., Peskin, C.~S., Oster, G.~F., 5 1992. What drives the
  translocation of proteins? Proc. Natl Acad. Sci.USA 89, 3770.

\bibitem[{Suhonen and Linna(2016)}]{Suhonen16}
Suhonen, P.~M., Linna, R.~P., 2016. Chaperone-assisted translocation of
  flexible polymers in three dimensions. Physical Review E 93, 012406.

\bibitem[{Sun and Luo(2014)}]{Sun14}
Sun, L.-Z., Luo, M.-B., 2014. Langevin dynamics simulation on the translocation
  of polymer through $\alpha$-hemolysin pore. Journal of Physics: Condensed
  Matter 26~(41), 415101.

\bibitem[{Tomkiewicz et~al.(2007)Tomkiewicz, Nouwen, and
  Driessen}]{Tomkiewicz07}
Tomkiewicz, D., Nouwen, N., Driessen, A. J.~M., 2007. Pushing, pulling and
  trapping – modes of motor protein supported protein translocation.
  Federation of European Biochemical Societies Letters 581~(15), 2820--2828.

\bibitem[{Vollmer and de~Haan(2016)}]{16HaanJCP}
Vollmer, S.~C., de~Haan, H.~W., 2016. Translocation is a nonequilibrium process
  at all stages: Simulating the capture and translocation of a polymer by a
  nanopore. The Journal of chemical physics 145~(15), 154902.

\bibitem[{Wei-Ping~Cao and Luo(2015)}]{Cao15}
Wei-Ping~Cao, Q.-B.~R., Luo, M.-B., 2015. Translocation of polymers into
  crowded media with dynamic attractive nanoparticles. Physical Review E 92,
  012603.

\bibitem[{Yu and Luo(2011)}]{kaifujacs}
Yu, W., Luo, K., 2011. Chaperone-assisted translocation of a polymer through a
  nanopore. Journal of the American Chemical Society 133, 13565.

\bibitem[{Yu and Luo(2014)}]{kaifuPRE14}
Yu, W., Luo, K., 2014. Polymer translocation through a nanopore driven by
  binding particles: Influence of chain rigidity. Physical Review E 90, 042708.

\bibitem[{Zandi et~al.(2003)Zandi, Reguera, Rudnick, and Gelbart}]{Z&R}
Zandi, R., Reguera, D., Rudnick, J., Gelbart, W.~M., 7 2003. What drives the
  translocation of stiff chains? Proceedings of the National Academy of
  Sciences USA 100, 8649.

\end{thebibliography}

\end{document}